\documentclass{article}    

\usepackage{graphicx}

\title{\textbf{NuRules and Objective/Observer Measurement}}  
\author{Richard Mould\footnote{Department of Physics and Astronomy, State University of New York, Stony Brook,
\mbox{New York} 11794-3800, http://ms.cc.sunysb.edu/\~{}rmould}}  
\date{}    
   
\begin{document}             

\maketitle              

\begin{abstract}

New rules are proposed to govern the collapse of a wave function during measurement.  These rules apply with or without an
observer in the system.  They overcome an absurdity that was previous found when an objective state reduction is combined
with an observer-based state reduction.      
   
\end{abstract}

\section*{Introduction}
	The argument of a previous paper leads to an absurdity when an \emph{objective measurement} is combined in a certain way
with an \emph{observer measurement} \cite{RM1}.  This was taken as an indication that objective measurements do not exist
in \mbox{Nature - because} they conflict with empirically verifiable observer measurements.  It was concluded that the
rules of measurement must require the presence of a conscious observer.  The author had previously published a set of
rules (1-4) that meet this requirement \cite{RM2}.

	I subsequently discovered that rules (1-4) in ref.\ 2 could be modified in such a way as to overcome the difficulties in
ref.\ 1.  The modified rules are called \emph{nuRules (1-4)}.  The nuRules are the same as the original rules (in ref.\ 2),
except that the \emph{basis states of reduction} are extended to allow the possibility of a wave collapse in systems in
which there are no observers.  

NuRule (1) is identical with rule (1) in ref.\ 2.

\vspace{0.3cm}
\textbf{nuRule (1)}: \emph{For any subsystem of n components in an isolated system with a square modulus equal to s, the
probability per unit time of a stochastic choice of one of those components at time t is given by $(\Sigma_nJ_n)/s$, where
the net probability current $J_n$ going into the $n^{th}$ component at that time is positive.}
\vspace{0.3cm}

The difference between the old and new rules begins with nuRule (2).  This defines the new \emph{basis states} of a
reduction (called \emph{ready} states) and provides for their introduction into solutions of the Schr\"{o}dinger
equation. 

\vspace{0.3cm}
\textbf{nuRule (2)}: \emph{If the Hamiltonian gives rise to new components that are locally incoherent with, and are
discontinuous with the old components or with each other, then all states that are included in the new components will be
ready states.}
\vspace{0.3cm}

If a state is not a ready state it will be called a \emph{realized} state.  A realized brain state is understood to be a
\emph{conscious} brain state.  We therefore preserve the dual categories `conscious' and `ready' applied to brains in
ref.\ 2, and extend the dualism to all objects in the form of `realized' and `ready'.  

NuRule (3) is the same as rule (3) in that it provides for a collapse of a wave function, changing a stochastically chosen ready state to a realized state.    

\vspace{0.3cm}
\textbf{nuRule (3)}: \emph{If a component containing ready states is stochastically chosen, then all of the states in that
component will become realized, and all other components will be immediately reduced to zero.}
\vspace{0.3cm}

NuRule (4) is different from Rule (4) in that it applies to \emph{all} ready states, not just to ready brain states.

\vspace{0.3cm}
\textbf{nuRule (4)}: \emph{A transition between two components is forbidden if each is an entanglement containing a ready
state of the same object.}
\vspace{0.3cm}

It is nuRule (4) that resolves the difficulty found in the previous paper \mbox{(ref.\ 1)} by allowing state reduction
to occur across the board.

\section*{The Interaction}
Let a particle and detector interact.  The state of the system prior to interaction is given by 
\begin{displaymath}
\Phi(t) = exp(-iHt)\psi_iD_i
\end{displaymath}
where $\psi_i$ is the initial state of the particle, and $D_i$ is the initial state of the detector.  After interaction at
time $t_0$, the particle and detector will become entangled and evolve into two decoherent components
\begin{equation}
\Phi(t \ge t_0) = \psi(t)D_0 + \underline{D}_1(t)
\end{equation}
where the first is the partially scattered particle together with the detector in its ground state, and the second is the
detector in its excited or capture state.  The second component will be zero at $t_0$ and increase with time.  This
interaction will be called the `primary' interaction.

The underline $\underline{D}_1(t)$ in eq.\ 1 means that this detector is a ``ready" state as required by nuRule (2), as
opposed to $D_0$ which is a ``realized" state in this equation.  In my mind, a realized state has a higher ÔrealityÕ status
than a ready state, although this ontological distinction has no operational significance in this treatment.  

Similarly, an underlined brain state $\underline{B}$ is now understood to be a ready brain state, and an un-underlined or
realized brain state $B$ is a conscious brain state.  This reverses the underline convention in ref.\ 2.  

As in ref.\ 1, the interaction in eq.\ 1 can persist for an indefinite period of time, depending on the cross section of
the interaction and the length of time it takes for the particle to pass over the detector.  The system may not collapse at
all, corresponding to there being no capture.  As before, the possibility of capture and its timing is a function of the
probability current flowing into the second component from the first component.

\section*{Application of the NuRules }
As in ref.\ 1, we imagine that the approaching particle has a total probability of
capture equal to 60\%, and we let an observer look at the detector at a time when there is only a 50\% probability of
capture.  So the observer looks at the primary interaction before it is complete.  

Prior to that observation, the primary interaction is given by
\begin{equation}
\Phi(t_{ob} > t \ge t_0) = [\psi(t)D_0 + \underline{D}_1(t)]\otimes X
\end{equation}
where $X$ is the unknown brain state of the observer before the observer and the detector interact at $t_{ob}$.  Current
flows from the first to the second component inside the square bracket.  It is given that 50\% of the time there will be a
\mbox{nuRule (1)} stochastic choice of the ready state $\underline{D}_1(t)$, and that that will lead to a nuRule (3) state
reduction of eq.\ 2 at the  time $t_{sc(pr)}$.
\begin{equation}
\Phi(t_{ob} > t \ge t_{sc(pr)} > t_0) = D_1(t)\otimes X
\end{equation}
The observation begins sometime later, initiating the observer's physiological interaction
\begin{equation}
\Phi( t \ge t_{ob} > t_{sc(pr)} > t_0) = D_1(t)\otimes X + \underline{D}'_1(t)\underline{B}_1
\end{equation}
where the entangled component $\underline{D}'_1(t)\underline{B}_1$ is zero at $t_{ob}$ and increases with time.  The
nuRule (1) stochastic trigger then hits the second component of eq.\ 4 at time $t_{sc(ob)}$.
\begin{equation}
\Phi( t \ge t_{sc(ob)} > t_{ob} > t_{sc(pr)} > t_0) = D_1B_1
\end{equation}
This happens 50\% of the time.  So the observer becomes initially conscious of the detector in its capture state 50\% of
the time.  

The remaining 50\% of the time, the observer interaction begins \emph{before} there is a stochastic hit of any kind.  The
initial physiological interaction is therefore 
\begin{eqnarray}
\Phi(t \ge t_{ob} > t_0) &=& \psi(t)D_0 \otimes X + \underline{D}_1\otimes X\\
&+& \underline{\psi}'(t)\underline{D}_0\underline{B}_0 + \underline{D}_1'(t)\underline{B}_1\nonumber
\end{eqnarray}
The states in the second row are equal to zero at $t_0$ and increase with time.   According to nuRule (2), the second row
states are all ready states including the ready brain states $\underline{B}_0$ and $\underline{B}_1$; so the observer is
not yet conscious of the detector in either its ground state or its capture state.

	The effect of nuRule (4) is to \emph{eliminate the fourth component} in eq.\ 6.  Current cannot flow from the second to
the fourth component, or from the third to the fourth component, because they all contain ready states of the detector. 
NuRule (4) therefore removes the fourth component from the picture, thereby avoiding the absurdity found in ref.\ 1.   We
then have
\begin{eqnarray}
\Phi(t \ge t_{ob} > t_0) &=& \psi(t)D_0 \otimes X + \underline{D}_1\otimes X\\
&+& \underline{\psi}'(t)\underline{D}_0\underline{B}_0 \nonumber
\end{eqnarray}

In this equation, current will flow from the first component to the second and from the first to the third.  The total
time integral of flow into these components is equal to 1.0.  Since we assumed that the second component was not
stochastically chosen prior to $t_{ob}$, and because we will exclude the possibility of a particle capture \emph{during}
the observer's physiological interaction, a stochastic hit on the third component at a time $t_{sc(ob)} > t_{ob}$ is a
certainty.  This will give the reduction 
\begin{displaymath}
\Phi(t = t_{sc(ob)} > t_{ob} > t_0) = \psi(t)D_0B_0
\end{displaymath}
which means that the other 50\% of the time the observer will initially see the detector in its ground state, so there
hasn't yet been a particle capture.  The interaction will continue giving
\begin{equation}
\Phi( t \ge t_{sc(ob)} > t_{ob} > t_0) = \psi(t)D_0B_0+ \underline{D}_1(t)\underline{B}_1
\end{equation}
where the second component is zero at $t_{sc(ob)}$ and increases with time.

According to the initially given conditions, there is still a 10\% possibility of a subsequent stochastic hit on that
component, or an over-all probability of capture equal to 60\%.  If there is no capture in the remaining time, then eq.\ 8
will be the final result, leaving $\underline{D}_1(t)\underline{B}_1$ a dormant phantom component that will serve no
further useful purpose\footnote{If a second observer looks at eq.\ 8, the phantom component will not result in a
measurement of any kind because nuRule (4) will forbid current flow out of the phantom.  See appendix.} .  However, if the
particle is captured during the remaining time that the observer looks at the detector, then there will be another
stochastic hit at time $t_{sc(f)}$ due to the primary interaction current flowing into the second component in eq.\ 8. 
This will give
\begin{equation}
\Phi( t \ge t_{sc(f)} > t_{sc(ob)} > t_{ob} > t_0) = D_1B_1 
\end{equation}

The \textbf{net result} of these nuRules is that when the observer first looks at the detector, he will experience it in
its capture state 50\% of the time (eq.\ 5), and in its ground state 50\% of the time (eq.\ 8).  However, in the latter
case, there may still be a particle capture during the time that remains for the primary interaction.  If that happens, the
observer will see the detector in its capture state (eq.\ 9).  

There is no observational absurdity as was found in ref.\ 1.  This is avoided because nuRule (4) eliminates the fourth
component in eq.\ 6.

\section*{Significance of NuRule (4)}
	Common sense would expect the first three nuRules to appear in any collection of rules that govern measurement in quantum
mechanics.  The first one provides for the existence of a stochastic trigger, the second provides for the definition and
introduction of the basis states of reduction, and the third provides for the collapse itself.  However, the fourth nuRule
is unexpected.  It was originally introduced to avoid an anomaly when the (old) rules were applied to a situation
involving two observers (in ref.\ 2 see section: Two Observers - Rule 4).  A similar anomaly appears in the
appendix of the present paper.  In addition, the original rule (4) was used repeatedly to good effect in an earlier paper
involving Schr\"{o}dingerÕs cat \cite{RM3} as is explained in ref.\ 2.

	A physical significance of the fourth rule (or nuRule) began to emerge in the Schr\"{o}dinger cat paper (ref.\ 3), but it
is more clearly expressed in another paper involving the observation of a $\beta$-counter \cite{RM4}.  It is shown
there that a conscious observer subject to the rules (old or new) is fully and continuously integrated into the observed
system.  However, that will be true only when the fourth rule is included. Without this rule, an observer might miss one
or more of the scale readings on the counter, for skipping over states is a natural consequence of second order
transitions quantum mechanics.  But with this rule or nuRule in effect, `ready' second order transitions are not
possible.  So each reading 0, 1, 2, 3, etc. on the face of the counter will be witnessed sequentially by a conscious
observer.  

	Another interesting application of nuRule (4) concerns \emph{florescent pulsing} in the case of the 3-level atom.  If the
transition to one excited state of the atom is very strong relative to the other, then it is known experimentally that the
atom will alternate from one radiation decay mode to the other \cite{HD}.  It will radiate repeatedly at the strong
frequency for a time, then it will become dark for a time corresponding to its taking a long time to decay in the weak
mode.  It is common to appeal to the notion of a ``null" measurement to explain the dark period.  But there is no such
thing as a null measurement in this treatment.  Instead, nuRule (4) explains the onset and persistence of the dark period
in a florescent pulse
\cite{RM5}.

\section*{Appendix}
Suppose a second observer is standing by while the first observer consciously interacts with the detector prior to a
capture.  The state is then
\begin{equation}
\Phi( t \ge t_{sc(ob)} > t_{ob} > t_0) = [\psi(t)D_0B_0 + \underline{D}_1(t)\underline{B}_1]\otimes X 
\end{equation}
where the expression in the brackets is eq.\ 8, and $X$ is an unknown state of the second observer prior to his interacting
with the system.  When a product of brain states appears in the form $BB$ or $B\otimes X$, the first term will refer to
the first observer and the second to the second observer.  If the second observer interacts with the detector at
$t_{ob2}$ before the interaction is over, his physiological interaction according to nuRule (2) will be 
\begin{eqnarray}
\Phi(t \ge t_{ob2} > t_{sc(ob)} > t_{ob} > t_0) &=& \psi(t)D_0B_0\otimes X + \underline{D}_1(t)\underline{B}_1\otimes  X\\
&+& \underline{\psi}'(t)\underline{D}_0\underline{B}_0\underline{B}_0 +
\underline{D}'_1(t)\underline{B}_1\underline{B}_1\nonumber
\end{eqnarray}
where the second row is zero at $t_{ob2}$, and increases with time.  NuRule (4) forbids current flow into the fourth
component, so the interaction is really
\begin{eqnarray}
\Phi(t \ge t_{ob2} > t_{sc(ob)} > t_{ob} > t_0) &=& \psi(t)D_0B_0\otimes X + \underline{D}_1(t)\underline{B}_1\otimes 
X\nonumber\\ &+& \underline{\psi}'(t)\underline{D}_0\underline{B}_0\underline{B}_0 \nonumber
\end{eqnarray}
The integral of current flow to the second and third components is equal to 1.0.  Since we assumed that the second
component was not stochastically chosen prior to $t_{ob2}$, and because we will exclude the possibility of a particle
capture \emph{during} the second observer's physiological interaction, a stochastic hit on the third component at a time
$t_{sc(ob2)} > t_{ob2}$ is a certainty.  This gives the reduction
\begin{displaymath}
\Phi( t = t_{sc(ob2)} > t_{ob2} > t_{sc(ob)} > t_{ob} > t_0) = \psi(t)D_0B_0B_0 
\end{displaymath}
Both observers are here conscious of the detector in its ground state, which means that the  second observer agrees with
the first that the particle has not yet been captured.  Since the interaction is not yet over, it will continue in the form
\begin{displaymath}
\Phi( t \ge t_{sc(ob2)} > t_{ob2} > t_{sc(ob)} > t_{ob} > t_0) = \psi(t)D_0B_0B_0 +
\underline{D}_1(t)\underline{B}_1\underline{B}_1 
\end{displaymath}
where the second component is zero at $t_{sc(ob2)}$ and increases with time.  This means that there is still the
possibility that both observers will experience a capture.  

	NuRule (4) therefore spares us another anomalous result.  If the fourth component in eq.\ 11 had not been eliminated,
there would have been a finite probability that both observers would witness a capture at time $t_{ob2}$, simply because
the second observer makes his observation at that time.  That of course is absurd.  The phantom component in eqs.\ 8 and
11 played to part in this result.


\begin{thebibliography}{99}

\bibitem{RM1}R.A. Mould, ``The Case Against Objective Measurement", 

quant-ph/0309048


\bibitem{RM2}R.A. Mould, ``Quantum Brain States", \emph{Found. Phys.} \textbf{33}(4) 591-612 (2003),
quant-ph/0303064


\bibitem{RM3}R.A. Mould, ``Schr\"{o}dinger's Cat: The rules of engagement",
 
quant-ph/0206065

\bibitem{RM4}R.A. Mould, ``Rule (4) and Continuous Observation", quant-ph/0309029

\bibitem{HD}H. Dehmelt, \emph{Bull. Am. Phys. Soc.}, \textbf{20}, 60 (1975)

\bibitem{RM5}R.A. Mould, ``NuRule (4) and the Three Level Atom", quant-ph/0309125

\end{thebibliography}
\end{document}